\documentclass{PoS}

\bibliography{/path/to/your/bibliographic/file}
\title{Clear anti-correlation between luminosity and high energy cutoff in
the low/hard state of the black hole candidate GX339--4}

\ShortTitle{Clear anti-correlation between luminosity and high energy
cutoff}

\author{\speaker{Takehiro G. Miyakawa}, Kazutaka Yamaoka, Atsumasa Yoshida, Koji Saito,\\
        Department of Physics and Mathematics, Aoyama Gakuin
	University, Fuchinobe 5-10-1, Kanagawa 229-8558, Japan\\ 
        E-mail: \email{tmiyakawa@phys.aoyama.ac.jp, yamaoka@phys.aoyama.ac.jp}}

\author{Tadayasu Dotani, and Hajime Inoue\\
        The Institute of Space and Astronautical Science,
         Japan Aerospace Exporlation Agency, Yoshinodai 3-1-1,
	 Sagamihara, Kanagawa 229-8510, Japan \\}

\abstract{We have analyzed the 171 RXTE data sets of the black hole
candidate GX 339-4 in the low/hard state during its 1996--2005
outburst. All the broadband spectra were successfully modeled by a
simple analytic model, power-law with an exponential cutoff. The
obtained energy cutoff($E_{\rm{cut}}$) is distributed over 50--300
keV, and the photon index over 1.4--1.6. We found a clear correlation
($E_{\rm{cut}}$ is proportional to $L^{-0.75 \pm 0.04}$) between
luminosity in 2--200 keV (L) and $E_{\rm{cut}}$ when L is larger
than 5$\times 10^{37}$ erg $s^{-1}$, while $E_{\rm{cut}}$ is almost
constant at 200 keV when L is smaller than 5$\times 10^{37}$ erg
$s^{-1}$. This anti-correlation is unchanged by adopting the more physical model of
thermal Comptonization by Sunyaev and Titarchuk, although a slightly
different relation is obtained as the electron temperature is
proportional to $L^{-0.23 \pm 0.02}$. These anti-correlations are
qualitatively explained by a picture where the energy flow rate from protons
to electrons balance with cooling due to inverse Compton scattering. }  

\FullConference{VI Microquasar Workshop: Microquasars and Beyond\\
		 September 18-22, 2006\\
		 Como, Italy}

\begin{document}
\section{Introduction}
Energy spectra of the black hole candidates (BHCs) have mainly
two spectral states: high/soft state and low/hard state. 
It has been established that the high/soft state happens at
comparatively high mass accretion rate, and there is an optically thick
and geometrically thin accretion disk which extends to three
Schwarzschild radii from the central black hole. However, the origin of the 
low/hard state has not been clarified yet. 
X--ray spectra in the low/hard states  are well represented by 
a power law with a photon index of 1.4$\sim$1.7 (Tanaka $\&$ Shibazaki 1996), 
The energy cutoff of the spectrum ($E_{\rm{cut}}$) is seen at $\sim$100 keV by 
past gamma-ray observations (Grove et al. 1998). 
It is believed that there is a high temperature corona inside the standard disk, and
radiation is produced by inverse Compton scattering from a part of soft
photons via optically thin plasma with very high temperature around the
black hole. 
Indeed, the broadband spectra of many black hole binaries in the
hard state are successfully modeled by thermal Comptonization model
(Dove et al. 1996, Pountanen $\&$ Svensson 1996), although location and
geometry of the corona are still debated. \\
 GX 339--4 was found by the X--ray satellite OSO--7 in 1971 (Markert et
al. 1973). Since it was similar to Cygnus X--1 in terms of the feature of
spectrum and short time variations, it was suggested as a black hole
candidate (Samimi et al. 1979). This source is one of the best studied
BHCs at X rays and gamma rays by various instruments: Ginga/LAC (Ueda,
Ebisawa and Done 1994), CGRO/OSSE(Grabelsky et al. 1995, Smith et
al. 1996), ASCA (Wilms et al. 1998), RXTE (Smith et al. 1996), and
Beppo--SAX (Corongiu et al. 2003). GX 339--4 exihited the five different
spectral states in the previous outbursts (Tanaka $\&$ Shibazaki 1996),
but often stayed in the low/hard state. All the spectra in the low/hard
state are roughly explained by a power law with an exponential cutoff or
thermal Comptonization model, requiring some modification of reflection
component, soft excess and iron-K lines. Detailed broadband analysis in
2-1000 keV was done using the Ginga, RXTE and OSSE data in 1991 and 1996
observations by Zdziarski et al. (1999) and Wardzi$\acute{n}$ski et
al. (2002). Wardzi$\acute{n}$ski et al. (2002) concluded that the four
spectra in the hard state have very similar intrinsic X-ray slope of photon
index $\cong$1.75. On the other hand, they found a possible correlation that
the high-energy cutoff energy decreases with luminosity
increasing. Zdziarski et al. (2004) studied the long term behaviour
using 16-years light curves by compiling the GINGA/ASM, CGRO/BATSE, and
RXTE/ASM data, and also found that the electron temperature depends on
the luminosity by showing a correlation between the BATSE flux and
photon index in the 20--160 keV range.  

\begin{figure}
\includegraphics[width=.32\textwidth]{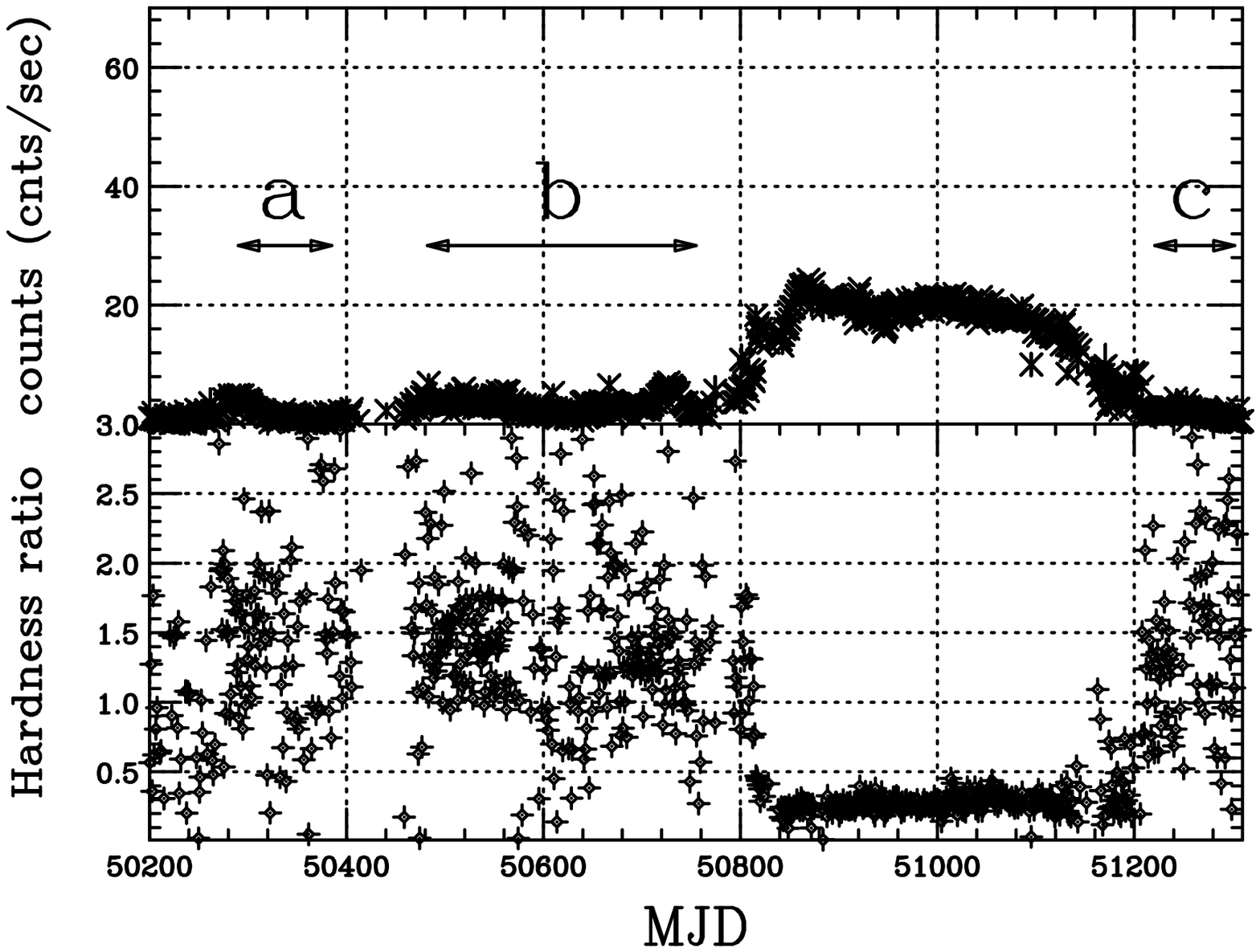}
\hspace{0.0cm}
\includegraphics[width=.32\textwidth]{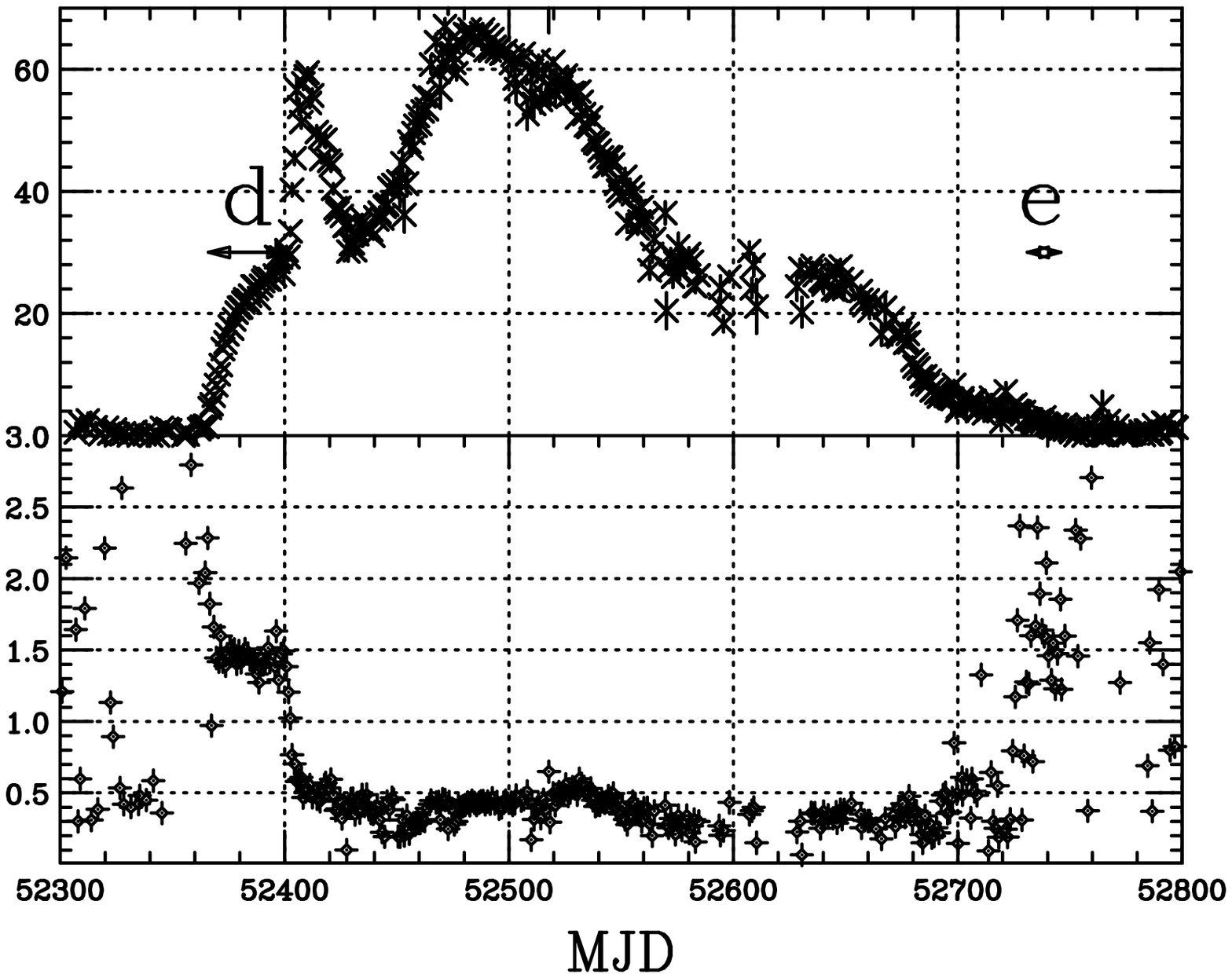}
\hspace{0.0cm}
\includegraphics[width=.32\textwidth]{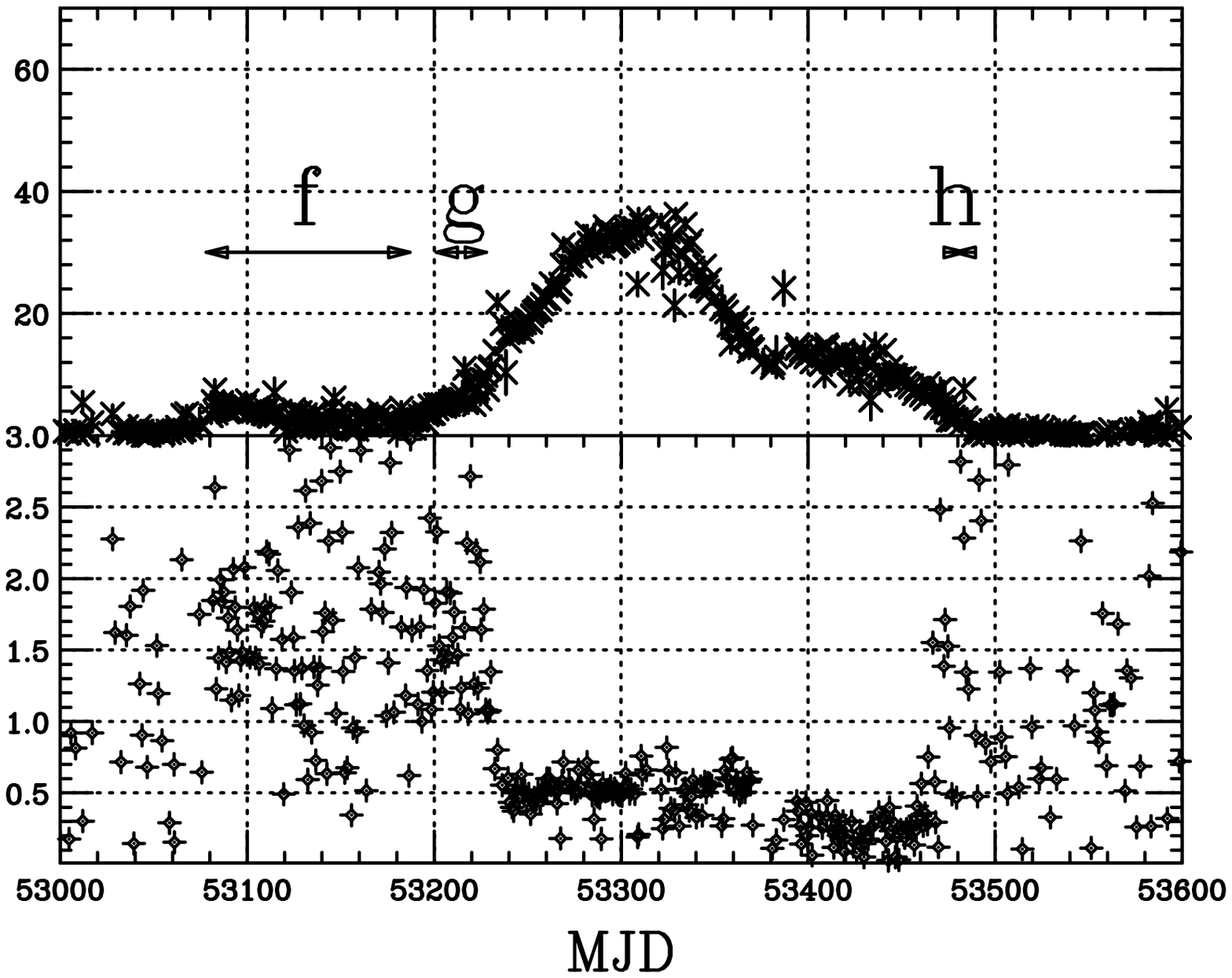}
\caption{RXTE/ASM light curve and hardness ratio of GX 339--4 from 1996 to 2005. }
\label{lc_1996_2005}
\end{figure}

As mentioned above, many authors have studied some observations during
individual outbursts of this source. Hence, the correlation among various
parameters (high energy cutoff, photon index, and so on) are poorly
understood. Thus, in order to clarify the radiation mechanism of the
low/hard state, we first performed a systematic study of detailed
correlations among spectral parameters using the large public RXTE data
archives.  \\
In this paper, we report on the results of the analysis of 171 RXTE data sets 
of GX 339--4 in the low/hard state during its 1996-2005 outburst. In
Section 2, we describe the procedure of analysis. In section 3, we
present the results and discussions. 

\section{Observations and Data Reduction}
GX 339--4 has been observed over $\sim$10 years by RXTE satellite. 
Figure $\ref{lc_1996_2005}$ shows the 1.5--12 keV RXTE/ASM light
curve. GX 339--4 underwent three outbursts in 1997, 2002, and 2004 during 
the past 10 years. According to McClintock $\&$ Remillard (2004), it is known 
 that HR2 in the low/hard state shows $\sim$ 1.5 where HR2 is the 
ratio of ASM counts in 5--12 keV to counts in 3--5 keV. Base on this 
criterion, we selected the data during the hard state from the publicly 
available RXTE data archive at NASA/HEASARC. Thus, we obtained 171 RXTE
data sets indicated by arrows in Figure$\ref{lc_1996_2005}$. 
For the RXTE data reduction, we used the publicly available software 
HEADAS 6.0.2 provided by NASA/GSFC. 

\begin{figure}
\includegraphics[width=.5\textwidth]{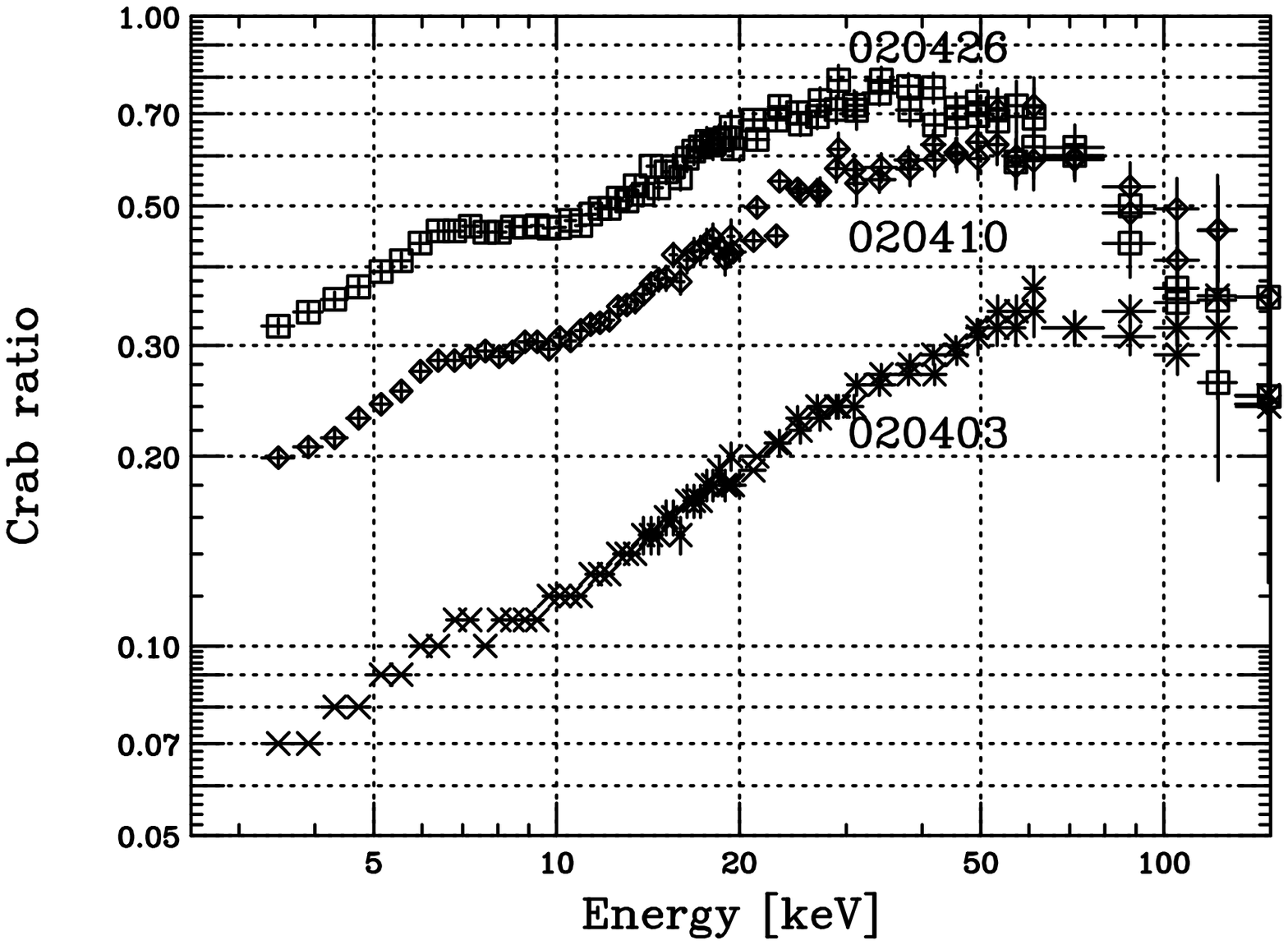}
\hspace{0.0cm}
\includegraphics[width=.5\textwidth]{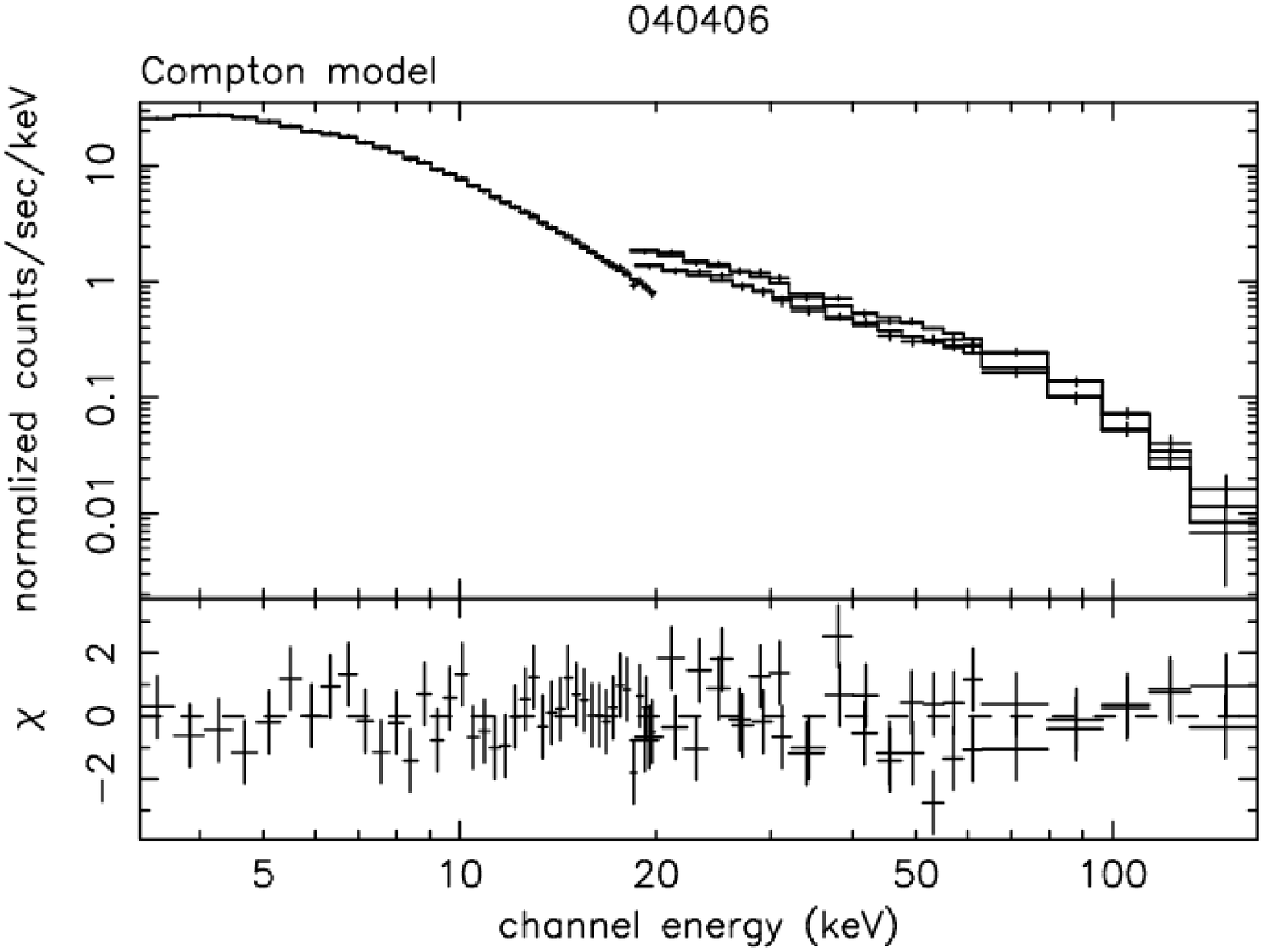}
\caption{Left: PHA ratio of GX 339--4 to Crab taken during different
 epochs. Right: The results of spectrum analysis observed in 6th April
 2004. (Low energy side is PCA. The high energy side is clusterA and
 clusterB. This analysis was done with the Compton model.)}
\label{ratio_and_Compton_spectrum}
\end{figure}

We used standard 2 mode data which has a time resolution of 16 sec and
129 energy channels covering the full range of the PCA detectors. The
PCA background was estimated with the background model for bright
sources. As for the HEXTE, we used the archive mode data with a time
resolution of 16 sec taken from both cluster A and B, and subtracted the
background taken from the rocking motion. 
Two spectral models are applied to all the data in carrying out the 
systematic spectral analysis of the continuum. One is a simple analytic model, 
power-law with an exponential cutoff (cutoff power-law), and the other is 
thermal Comptonization model introduced by Sunyaev $\&$ Titarchuk (1980) (compst in 
XSPEC). The cutoff power-law model consists of two parameters: a photon
index ($\alpha$) and an energy of spectrum cutoff ($E_{\rm{cut}}$).
The compst model also has two parameters: an electron temperature
(k$T_{\rm{e}}$) and a Thomson optical depth ($\tau$) of
 high temperature plasma.  These continuum models are further modified by galactic 
absorption (wabs in XSPEC) and smeared edge model (Ebisawa et al. 1994). 
The hydrogen column density was fixed to 5 $\times 10^{21}$cm$^{-2}$
 (Ilovaisky et al. 1986). The energy of the absorption edge and width in
 the smeared edge model is fixed at 7.11 keV due to the neutral iron-K
 edge and 10 keV, respectively. 
We added 2 $\%$ systematic errors to each PCA spectral bin so as to
${{\chi}_{\nu}}^2 \sim 1$ for the Crab nebula. The PCA energy range is
limited to 3$\sim$20 keV bacause HEXTE's effective area is larger than
that of PCA above 20 keV. We fit the data of PCA and HEXTE in 3--200
keV. simultaneously, leaving the relative normalization of PCA and HEXTE to
be free. The HEXTE normalization factor is found to be always
0.90--0.95. The current PCA and HEXTE response matrices give a consistent value of
photon index with 2.08$\pm$0.02 in 3-20 keV and 2.10$\pm$0.02 in 20-200
keV for the Crab Nebula, which is applicable to the joint fitting
between the two instruments. An X-ray luminosity (L) was calculated
based on the PCA flux over 2-200 keV assuming the distance to GX 339--4 
at 8 kpc (Zdziarski et al 2004). 
The obtained luminosity ranges from 7$\times 10^{36}$ erg s$^{-1}$ to
2.1$\times 10^{38}$ erg s$^{-1}$, corresponding to 2-20 $\%$ of the
Eddington limit ($L_{\rm{E}}$) assuming the black hole mass at 10$M_{\odot}$. 

\begin{figure}
\includegraphics[width=.5\textwidth]{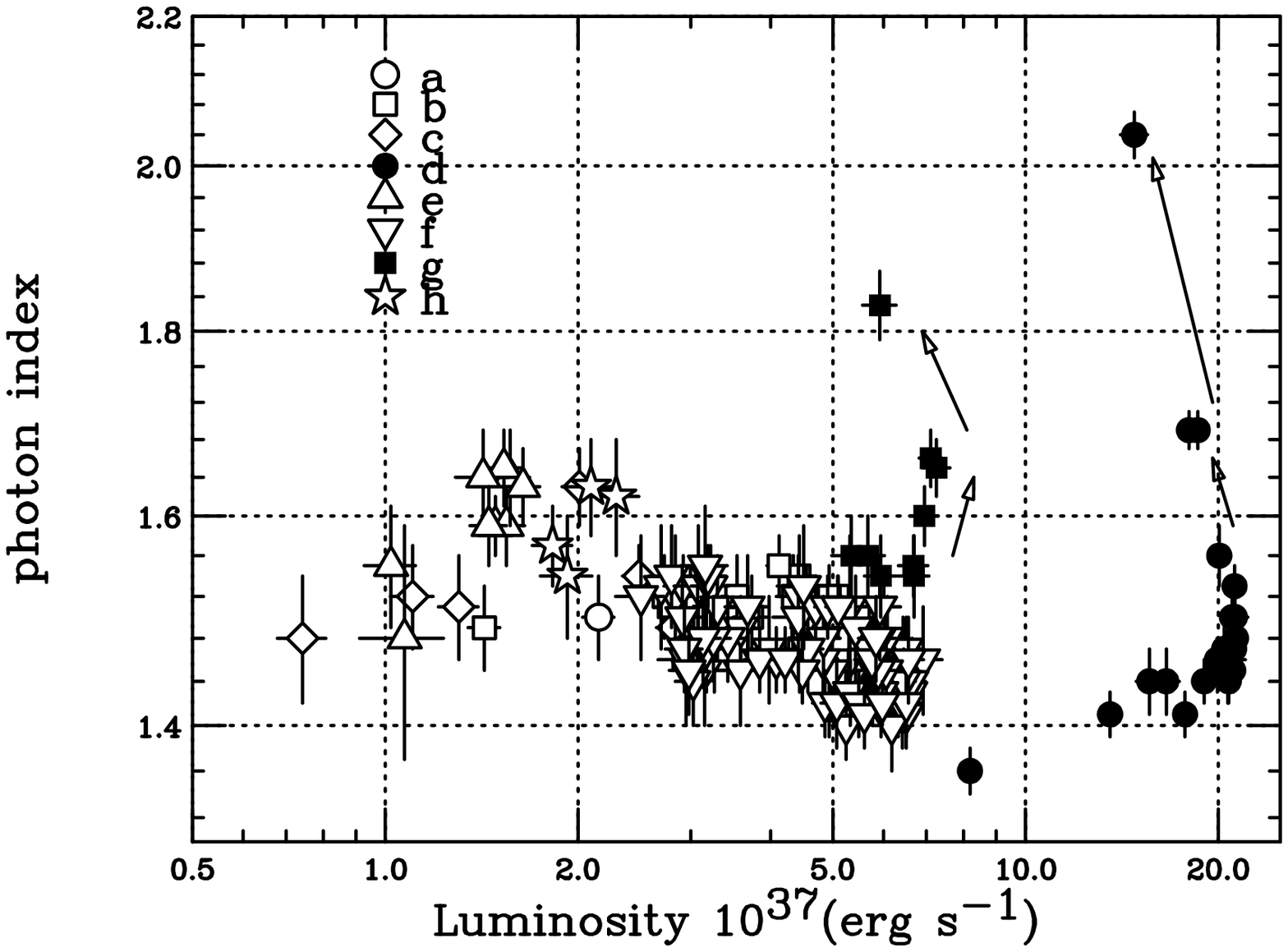}
\hspace{0.0cm}
\includegraphics[width=.5\textwidth]{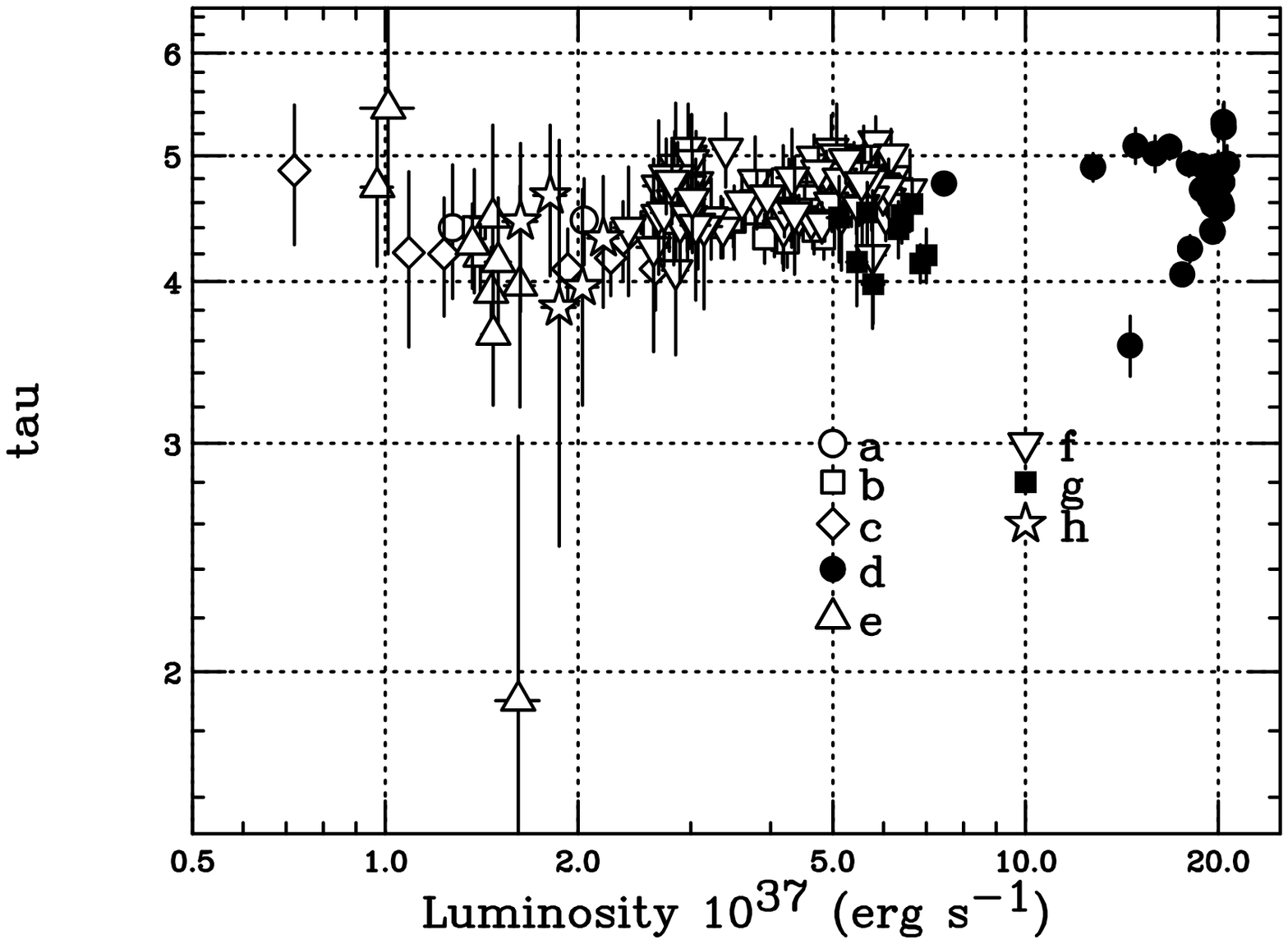}
\caption{Left: relation between luminosity in 2-200 keV and photon index. Right:
 relation between luminosity and optical depth}
\label{L_pho_tau}
\end{figure}

\section{Results and discussions}
\begin{figure}
\includegraphics[width=.5\textwidth]{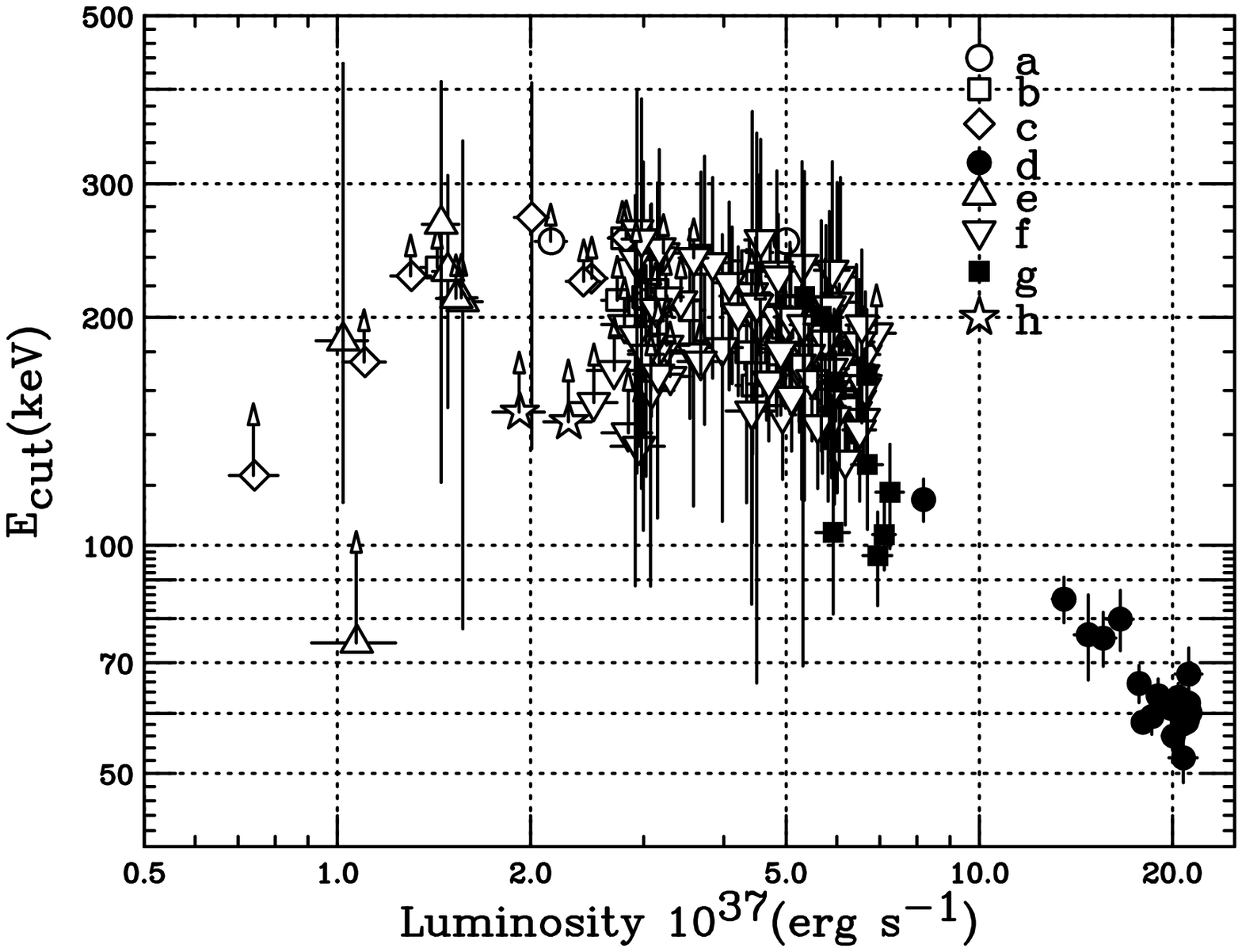}
\hspace{0.0cm}
\includegraphics[width=.5\textwidth]{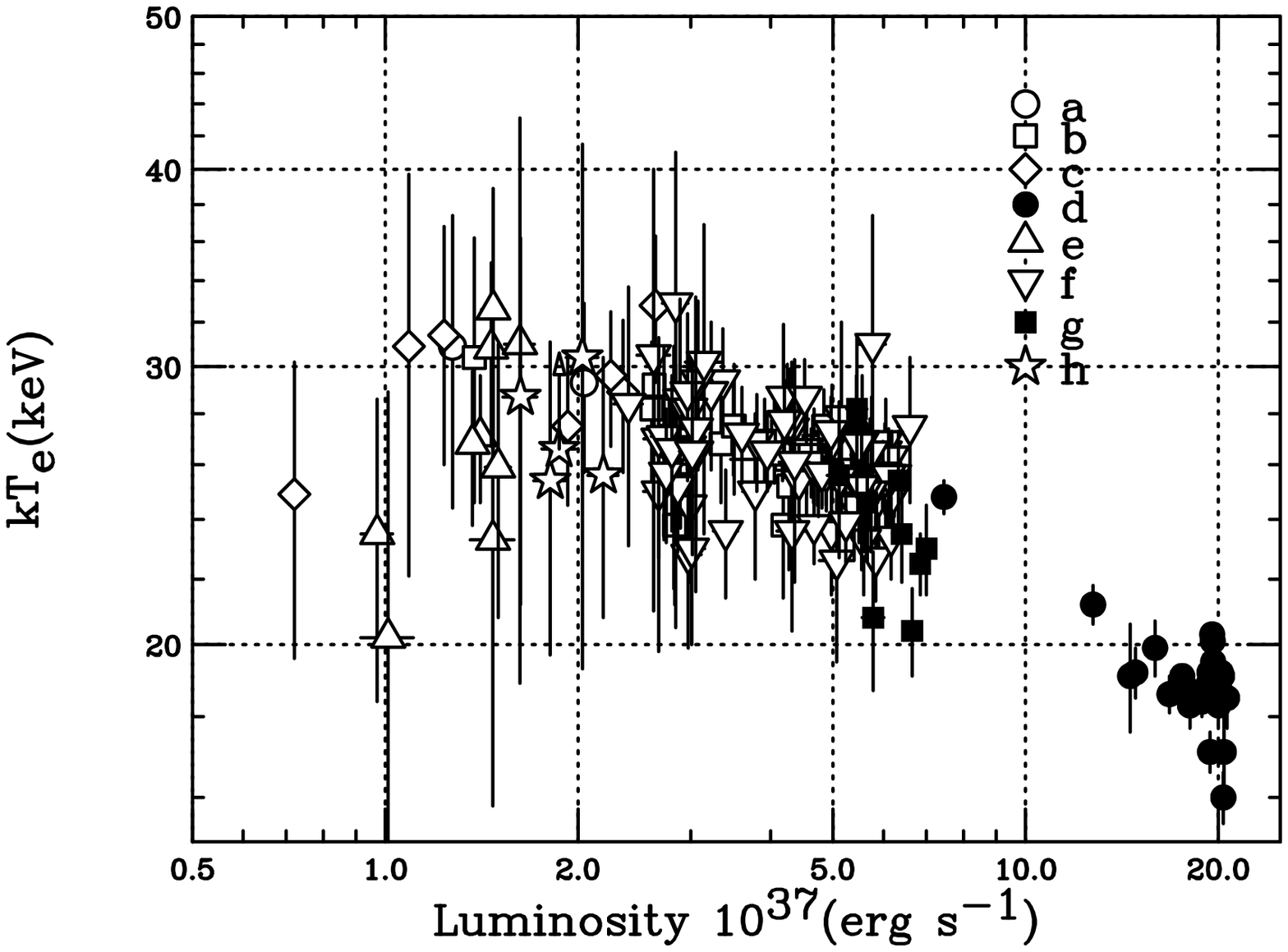}
\caption{Left: Relation between luminosity in 2-200 keV and energy cutoff
 ($E_{\rm{cut}}$). The results are for a fit with a cutoff power law
 model. Right: Relation between the electron temperature (${kT}_{\rm{e}}$) and
 luminosity in 2-200 keV. The results are from the COMPST model. }
\label{L_E_kT}
\end{figure}

In order to ascertain the presence of a high energy cutoff in a 
model-independent way, we first calculated the PHA ratio of GX 339-4 
to the Crab Nebula.  The Crab spectrum was extracted in the same way 
as above from data observed on April 28th 2002. 
A Crab spectrum in 3--200 keV was represented by a simple featureless 
power-law model with a photon index of 2.10$\pm$0.01. 
The left side of Figure $\ref{ratio_and_Compton_spectrum}$ shows 
the spectrum ratio of GX 339--4 to Crab. 
The spectra are from April 3th 2002 (MJD 52367),
April 10th 2002 (MJD 52374), and April 26th 2002 (MJD 52390), respectively.
As can be seen from this figure, we found a clear high energy cutoff at 
$\sim$30 keV, $\sim$50 keV, and $\sim$70 keV from upper to lower, suggesting 
that the peak energy varies with the X-ray luminosity, while 
the spectral slope, i.e. photon index, does not seem to vary significantly. \\
The right panel of Figure $\ref{ratio_and_Compton_spectrum}$ 
is sample in analyzing the spectrum taken on April 6 2004. 
Both models can represent the broadband spectrum over 3--200 keV 
 successfully. \\
The left panels of Figure $\ref{L_pho_tau}$ and Figure $\ref{L_E_kT}$ 
show the relation between parameters of the cutoff power law model. 
The points correspond from the different outbursts. 
The high energy cutoff($E_{\rm{cut}}$) ranges from 50 keV to 300 keV or
more where we cannot constrain the parameters well. 
As you can see from this figure, we found a clear anti-correlation between 
$L$ and $E_{\rm cut}$ when L$ > 5 \times 10^{37}$ erg s$^{-1}$. 
This relation follows an equation: $E_{\rm cut} \propto L^{-0.75 \pm 0.04}$. 
Errors quoted are at the 90$\%$ confidence level.
On the other hand, $E_{\rm{cut}}$ seems to be almost constant at 200 keV when 
L$ < 5 \times 10^{37}$ erg s$^{-1}$. 
Furthermore, we found that the epoch when the X--ray luminosity is larger than
  7$\times 10^{37}$ erg s$^{-1}$ corresponds to an initial rising phase
  of the outburst in 2002 and 2004. Unlike the high energy cutoff, the
  photon index is not sensitive to the X--ray luminosity. It is
  distributed over 1.4--1.7, which is consistent with the typical value of
  black hole candidates in the low/hard state. When the source becomes
  brighter than a certain value in the initial phase, the photon index
  shows 1.8--2.1. This may be considered as a part of state transitions
  from hard state to intermediate or very high state. \\
 In the right panel of Figure $\ref{L_pho_tau}$ and Figure $\ref{L_E_kT}$, 
we further show correlations between the parameters when we fit the
spectra with a more physical component: 
the thermal Compton model of Sunyaev $\&$ Titarchck. k$T_{\rm e}$ is
distributed over 20--30 keV, while the optical depth $\tau$ is distributed over 4$\sim$5.  
We also found a clear anti-correlation between of k$T_{\rm e}$ and $L$ of 
$kT_{\rm{e}} \propto L^{-0.23 \pm 0.02}$ when  L$ > 5 \times 10^{37}$ 
erg s$^{-1}$. The k$T_{\rm e}$ remains at 26 keV 
when L$ < 5 \times 10^{37}$ erg s$^{-1}$. \\

Next, We will now try to explain the anti-correlation between $L$ and 
k$T_{\rm e}$ quantitatively. Let us assume a high temperature corona
with a spherical geometry (radius R). The proton temperature (k$T_{\rm p}$) is assumed to be 
approximately constant at $\sim \frac{GM m_{\rm p}}{R}$ ($m_{\rm p}$ is
the proton mass and R is the radius), that is, the energy loss rate of protons is much smaller than viscous
heating rate through the accretion. 
The protons will give their energy to electrons through two-body
collisions and the energy loss rate per unit volume is given as
$\frac{\frac{3}{2}nkT_{\rm p}}{t_{\rm{pe}}}$ if $T_{\rm{p}}$
is much larger than $T_{\rm{e}}$ where $\rm{n}$ is a number density 
of plasma. $t_{\rm{pe}}$ ($\propto \frac{
{(\frac{kT_{\rm{p}}}{m_{\rm{p}}} + \frac{k T_{\rm{e}}}{m_{\rm{e}}})}^{\frac{3}{2}}}{n}$)
is the equipartition time-scale due to Coloumb collision (Spizter 1962). 
The cooling rate per unit volume is approximately given as 
$\frac{4k T_{\rm{e}}}{m_{\rm{e}} c^2}$$U_{\rm{rad}}$n${\sigma}_{\rm T}$ where
$c$ is the light speed, $U_{\rm{rad}}$ is the photon
flux density which is given by $U_{rad} \cong \frac{L \tau}{\pi R^2}$, 
and $\sigma_{\rm T}$ is the cross section of Thomson scattering.  
If $\frac{k T_{\rm{e}}}{m_e}$ is larger than $\frac{k T_p}{m_p}$, 
we will get $t_{pe} \propto \frac{T_{\rm{e}}^{\frac{3}{2}}}{n}$.
In the steady state, the energy flow rate from proton to electron 
should balance with the cooling rate due to inverse Compton scattering, 
i.e.  $\frac{\frac{3}{2} nk T_{\rm{p}}}{t_{\rm{pe}}} = \frac{4kT_{\rm{e}}}{m_{\rm{e}} c^2} U_{\rm{rad}} n {\sigma}_{\rm T}$. Using
these equations, we can work out the anti-correlation of $k T_{\rm{e}} \propto L^{-\frac{2}{5}}$. 
Thus, the obtained anti-correlation is quantitatively explained by the fact
that the radiation mechanism is due to inverse Compton scattering. 
Zdziarski et al.\ (1998) also suggest that the high energy cutoff
depends on the luminosity from a theoretical point of view of thermal
Comptonization. They showed a relation of $k T_{\rm{e}} \propto L^{-2/7}$ and
$k T_{\rm{e}} \propto L^{-1/6}$ in the advection dominant and cooling dominant 
case, respectively. In comparison with this prediction, our result,
$\sim$ --0.23, is well consistent with these values. 
Furthermore, the maximum luminosity in the low/hard state is predicted as 
 $L_{\rm max} \sim 0.15 y^{\frac{3}{5}} \alpha^{\frac{7}{5}} L_{\rm E}$
 in his model. The highest luminosity observed in 2002,
 $\sim$0.15$L_{\rm E}$, is quite agreement with this value.

\end{document}